\documentclass[twocolumn]{revtex4}
\def\address{\affiliation}
\usepackage[dvips]{graphicx}
\begin{document}

\title{
Charge transport of Bi$_{2-x}$Pb$_x$Sr$_2$ErCu$_2$O$_8$ single crystals 
near the insulator-superconductor transition
}

\author{
I. Terasaki*, T. Mizuno, K. Inagaki, and Y. Yoshino
}

\address{
Department of Applied Physics, Waseda University, Tokyo 169-8555, JAPAN
}

\begin{abstract}
We prepared a set of Pb-substituted Bi$_{2-x}$Pb$_x$Sr$_2$ErCu$_2$O$_8$ 
single crystals. 
The resistivity and the in-plane thermopower decrease
with increasing $x$ from 0 to 0.8, 
which indicates that the Pb substitution 
supplies holes in the CuO$_2$ plane.
For $x$=0.8, a tiny resistivity drop 
(a trace of superconductivity) is seen near 70 K, 
suggesting that the doping level is close to the critical concentration. 
The $c$-axis dielectric constant also systematically changes with $x$, 
and suggests that the charge inhomogeneity is suppressed by the Pb substitution.
\end{abstract}

\pacs{74.72.Hs, 77.22.Gm, 71.30.+h}
\keywords{
Bi-based cuprates, Dielectric loss and relaxation,
Metal-insulator transition\\
\\
$^*$Corresponding Author\\
Prof. I. Terasaki\\
Postal Address: Department of Applied Physics, Waseda University\\
3-4-1 Ohkubo, Shinjuku-ku, Tokyo 169-8555, Japan\\
Phone \& Fax: +81-3-5286-3116\\
Email: terra@mn.waseda.ac.jp
}

\maketitle

\section{Introduction}
As is well known, high-$T_c$ superconductivity is realized 
by doping carriers into the CuO$_2$ plane for the parent Mott insulator. 
The insulator-metal transition (IMT) 
of high-$T_c$ superconductors is significantly different 
from that of a doped semiconductor in the sense 
that the doped carriers exhibit superconductivity 
as soon as they become metallic. 
In fact, the mechanism of the IMT is quite unconventional. 
For La$_{2-x}$Sr$_x$CuO$_4$, for example, 
the doped carriers form a stripe, 
and the stripe direction changes 
from the diagonal to the parallel direction at IMT \cite{Wakimoto}.

The Bi-based cuprate Bi$_2$Sr$_2$CaCu$_2$O$_8$ 
(Bi-2212) is considered to be the counter part 
to La$_{2-x}$Sr$_x$CuO$_4$. 
This compound is away from the stripe instability, 
and no static stripe order is reported 
for the parent insulator of Bi-2212. 
We have studied the transport properties of the parent insulator 
of Bi-2212, and proposed that the high-temperature charge dynamics
is essentially similar to that for high-$T_c$ 
superconductors \cite{Kitajima,Takemura}. 

Previously we successfully prepared single crystals of 
the parent insulator Bi$_2$Sr$_2$RCu$_2$O$_8$ 
for various rare earth ions R
(R=Dy, Er, Y), and found that R=Er is most conducting \cite{Takemura}.
However the carrier concentration of the R=Er sample 
was 0.03 hole per Cu, which is still smaller than the critical
concentration ($\sim$0.05).
In order to dope more holes in the CuO$_2$ plane, 
we focus on the Pb substitution. 
As is well established, the Pb substitution for Bi 
supplies holes for the superconducting samples
of Bi$_2$Sr$_2$CuO$_6$ and Bi$_2$Sr$_2$CaCu$_2$O$_8$ \cite{Maeda,Motohashi}.
Here we report on the characterization and charge transport 
of Bi$_{2-x}$Pb$_x$Sr$_2$ErCu$_2$O$_8$ single crystals.

\begin{figure}
 \begin{center}
  \includegraphics[width=8cm,clip]{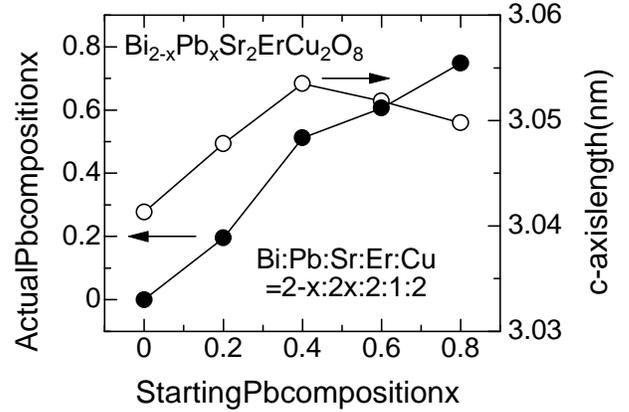}
 \end{center}
 \caption{the actual Pb content
 and the $c$-axis length of the prepared crystals 
 as a function of the starting Pb composition.}
 \label{fig1}
\end{figure}

\section{Experimental}
Single crystals of Bi$_{2-x}$Pb$_x$Sr$_2$ErCu$_2$O$_8$
were prepared by a Bi$_2$O$_3$/CuO flux technique.
A mixture of Bi$_2$O$_3$, PbO, SrCO$_3$, Er$_2$O$_3$ and CuO powder
was charged in an Al$_2$O$_3$ crucible,
heated up to 1373 K, slowly cooled down to 1173 K 
with a rate of 0.5-1 K/h. 
The ratio of the starting powder was Bi:Pb:Sr:Er:Cu=2-$x$:$2x$:2:1:2,
which is nearly stoichiometric except that the Pb content was doubled. 
Obtained crystals were plate-like, and the sizes were larger
for larger $x$ with a typical dimension of 3$\times$3$\times$0.05 mm$^3$
for $x$=0.6 and 0.8. 
An energy-dispersive x-ray (EDX) analysis revealed that
Bi+Pb and Sr are slightly excessive, and the content was estimated 
to be (Bi+Pb)$_{2.2}$Sr$_{2.7}$Er$_{0.7}$Cu$_2$O$_y$ 
(within an experimental error of 10\%). 

The resistivity was measured with the four-probe method,
and the $ab$-plane thermopower was measured with the steady-state
technique described previously \cite{Takemura}.
The $c$-axis dielectric constant from 10$^6$ to 10$^8$ Hz
was measured with an rf LCR meter (Agilent 4287A)
with a similar technique 
by B\"ohmer {\it et al.}\cite{Bohmer}.

\begin{figure}
 \begin{center}
  \includegraphics[width=7cm,clip]{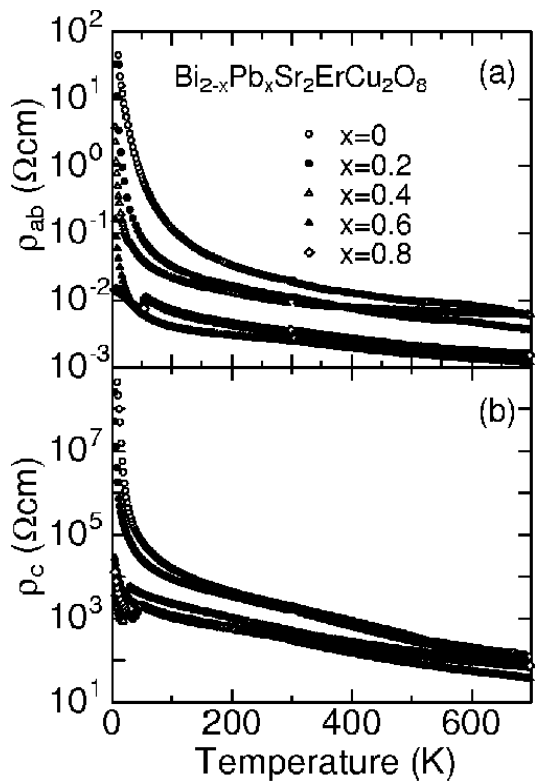}
 \end{center}
 \caption{Resistivity for
 Bi$_{2-x}$Pb$_x$Sr$_2$ErCu$_2$O$_8$ single crystals.}
 \label{fig3}
\end{figure}

\section{Results and Discussion}
In Fig. 1 is plotted the actual Pb content
and the $c$-axis length of the prepared crystals 
as a function of the starting Pb composition.
The actual Pb content is well controlled, and 
well corresponds to the starting composition. 
The $c$-axis length does not systematically change
with the Pb content, whose reason is not understood at present.

Figure 2 shows the $ab$-plane ($\rho_{ab}$) and 
$c$-axis ($\rho_c$) resistivity.
$\rho_{ab}$ and $\rho_c$ decrease with increasing Pb content, 
which indicates that the substituted Pb supplies 
holes in the CuO$_2$ plane.
For $x$=0.8, 
$\rho_{ab}$ at 300 K is of the order of m$\Omega$cm,
and a tiny drop (a trace of superconductivity) 
 in $\rho_{ab}$ and $\rho_c$ near 70 K is observed.
These results clearly indicate that
the doping level is very close to the IMT.

Figure 3 shows the $ab$-plane thermopower. 
Although the magnitude systematically decreases with the Pb content,
the increase in the hole concentration is much smaller than expected.
Estimated from the room-temperature thermopower \cite{Tallon},
the hole concentration per Cu is increased only from 0.03 to 0.04 
by increasing the Pb content from 0 to 0.8.
This is more than one order of magnitude smaller than the value
calculated by assuming that one Pb$^{2+}$ ion supplies one hole.
Perhaps this is because the substituted Pb removes the excess oxygen,
which compensates the carrier concentration.
In fact, the modulation structure in the Bi$_2$O$_2$ plane
due to the excess oxygen disappears near $x$=0.8 \cite{Tanaka}.

Figure 4 shows the $c$-axis dielectric constant at 80 K.
A systematic change with the Pb content is seen,
and the most conducting sample of $x$=0.8 exhibits
a small and frequency-independent value.
This is quite different from our previous measurement 
on the Pb-free Bi-2212 crystals, 
where the large magnitude
with the dielectric relaxation is seen \cite{yanagi}.
At present we cannot specify the reason of the difference,
but we wonder whether the excess oxygen plays an important role.
Recent scanning tunnel microscope/spectroscopy 
has revealed unexpected inhomogeneous electronic states of 
high-temperature superconductors. 
Pan {\it et al.} \cite{pan} found that the local density of states
are inhomogeneous in the CuO$_2$ plane of
the optimally doped Bi-2212,
where the doped carriers form a metallic patch.
Considering that the metallic patch is located near the excess oxygen,
we think it natural that the charge is more or less homogeneous
in the Pb substituted (and consequently excess-oxygen free) sample. 
Since the dielectric constant is a measure of the inhomogeneity \cite{yanagi},
it would be small where the charge is homogeneously distributed. 
In any case, the dielectric constant does not diverge near the IMT,
which makes a remarkable contrast 
with the IMT in doped semiconductors  \cite{rosenbaum}. 

\section{Summary}
Single crystals of the Pb-substituted Bi-2212 parent insulator 
Bi$_{2-x}$Pb$_x$Sr$_2$ErCu$_2$O$_8$
were successfully synthesised from $x$=0 to 0.8.
Holes are systematically doped with increasing Pb content,
but the increased concentration is as small as 0.01 per Cu
from $x$=0 to 0.8.
The most conducting sample of $x$=0.8 is 
very close to the IMT, and a tiny trace of 
superconductivity is observed.
The dielectric constant is measured to be small and 
frequency-independent, which would imply that 
the charge is  distributed 
more homogeneously than in the Pb-free sample.

%acknowledgments
The authors would like to thank K. Yoshida, S. Tajima, and S. Tanaka
for the technical support of rf impedance measurement.
They also thank A. Fujimori, N. Harima and K. Tanaka
for fruitful discussion.
This research was partially supported by the Ministry of Education, 
Culture, Sports, Science and Technology, 
Grant-in-Aid for Scientific Research (C),
2001, No. 13640374.

\begin{figure}
 \begin{center}
  \includegraphics[width=7cm,clip]{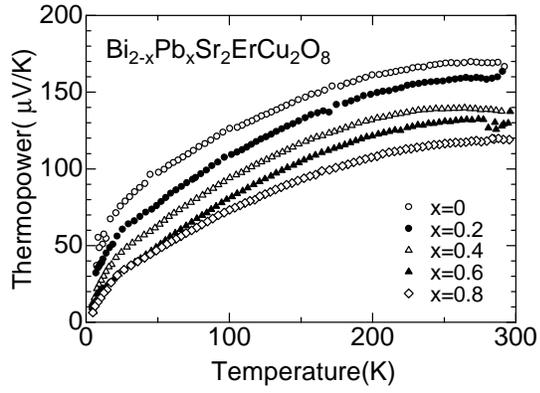}
 \end{center}
 \caption{$ab$-plane thermopower for 
 Bi$_{2-x}$Pb$_x$Sr$_2$ErCu$_2$O$_8$ single crystals.}
 \label{fig3}
\end{figure}

\begin{figure}
 \begin{center}
  \includegraphics[width=7cm,clip]{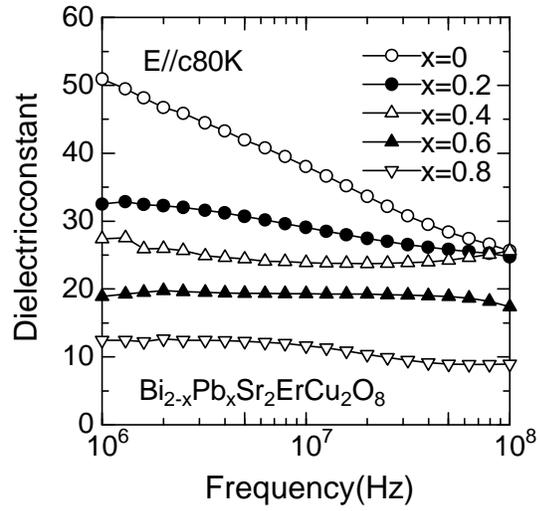}
 \end{center}
 \caption{$c$-axis dielectric constant for
 Bi$_{2-x}$Pb$_x$Sr$_2$ErCu$_2$O$_8$ single crystals.}
 \label{fig4}
\end{figure}

\end{document}